\documentclass[11pt,superscriptaddress,preprintnumbers,amsmath,amssymb,nofootinbib]{revtex4}
\usepackage{epsfig}  
\usepackage{graphicx}
\usepackage{hyperref}
\hypersetup{
    colorlinks=true,
    citecolor=red,
    linkcolor=blue,
    filecolor=green,      
    urlcolor=magenta,
}
\usepackage{color}
\usepackage{float}
\usepackage{amsfonts}
\usepackage{amsmath}
\usepackage{slashed}
\usepackage{soul}

\large

\begin{document}

\title{A unique discrimination between new physics scenarios in $b\rightarrow s\mu^+\mu^-$ anomalies}

\author{Ashutosh Kumar Alok}
\email{akalok@iitj.ac.in}
\affiliation{Indian Institute of Technology Jodhpur, Jodhpur 342037, India}

\author{Suman Kumbhakar~\footnote{Presently at Centre for High Energy Physics, Indian Institute of Science Bangalore 560012}}
\email{suman@phy.iitb.ac.in}
\affiliation{Indian Institute of Technology Bombay, Mumbai 400076, India}

\author{S. Uma Sankar}
\email{uma@phy.iitb.ac.in}
\affiliation{Indian Institute of Technology Bombay, Mumbai 400076, India}

\begin{abstract}
A number of observables related to the $b \to s \ell^+ \ell^-$ transition show deviations from their standard model predictions. 
A global fit to the current $b\rightarrow s\ell^+\ell^-$  data suggests several new physics solutions. Considering only one 
operator at a time and new physics only in the muon sector, it has been shown that the new physics scenarios 
(I) $C_9^{\rm NP}<0$, (II) $C_{9}^{\rm NP} = -C_{10}^{\rm NP}$, (III) $C_9^{\rm NP} = -C_9^{\prime \rm NP}$ 
can account for all data. In this work, we develop a procedure to  uniquely identify the correct new physics solution. The scenario II predicts a significantly lower value of $\mathcal{B}(B_s\to \mu^+\mu^-)$ and can be distinguished 
from the other two scenarios if the experimental uncertainty comes down by a factor of three. On the other hand, a precise measurement of the CP averaged angular observables $S_9$ in high $q^2$ bin of 
$B\to K^*\mu^+\mu^-$ decay can uniquely discriminate between the other two scenarios. 
We propose new methods, in terms of azimuthal angle asymmetries, to measure $S_9$ with 
the necessary precision.
\end{abstract}
 
\maketitle 

\section{Introduction} 

The quark level transition $b \to s\ell^+\ell^-$ ($\ell=e,\,\mu$) has immense potential to probe physics beyond Standard Model (SM). This decay is forbidden at the tree level within the SM and hence is highly suppressed. Further, the same quark level transition induces several decay modes such as $B \to X_s \ell^+\ell^-$, $B \to (K,\,K^*) \ell^+ \ell^-$,  $B_s \to \phi \ell^+ \ell^-$,  $B_s \to \ell^+ \ell^-$, thus providing a plethora of observables to probe new physics (NP). Due of these reasons, the $b \to s\ell^+\ell^-$  sector plays a pivotal role in  hunting physics beyond SM.  

The importance of this sector has increased considerably over last few years due to the fact that several deviations from the SM have been observed in decay modes induced by $b \to s\ell^+\ell^-$. These include measurements of the lepton flavor universality (LFU) violating ratios $R_{K}$ and $R_{K^*}$ \cite{rkstar,Rk2019,Abdesselam:2019wac,LHCb:2021trn}. The measured values of these observables  disagree with their SM predictions of $\approx$ 1 \cite{Hiller:2003js, Bordone:2016gaq} at the level of  $\sim 2.5-3\sigma$. This tension with the SM can be accounted by assuming new physics in $b\rightarrow s\, e^+\,e^-$ and/or $b\rightarrow s\mu^+\mu^-$~\footnote{A detailed study on the possibility of new physics in $b\to se^+e^-$ can be found in refs.~\cite{Kumar:2019qbv,Datta:2019zca,Alok:2020mvm}.}. Further, there are a few anomalous measurements which can be elucidated by considering new physics only in  $b \to s \mu^+ \mu^-$ transition. These include measurements of branching ratio of $B_s \to \phi\mu^+\mu^-$ \cite{bsphilhc2} and angular observable $P'_5$ in $B \to K^* \mu^+\mu^-$ decay \cite{Kstarlhcb1,Kstarlhcb2,Aaij:2020nrf}. The measured values disagree with the SM expectations at the  $\sim 4\sigma$ level \cite{sm-angular}. Hence one can account for all of these measurements simply by assuming new physics only in the muon sector.

This pile-up of anomalies in a coherent fashion can be considered as a signature of new physics (NP). This NP can be quantified in a model independent way, within the framework of effective field theory, by the addition of new operators to the SM effective Hamiltonian. Model independent analysis serves as a guideline for constructing specific new physics models which can account for these anomalies. In order to identify the Lorentz structure of possible new physics, several groups have performed global fits to all available data in the $ b\to s\mu^+ \mu^- $  sector \cite{Alguero:2019ptt,Alok:2019ufo,Ciuchini:2019usw,DAmico:2017mtc,Datta:2019zca,Aebischer:2019mlg,Kowalska:2019ley,Arbey:2019duh,Geng:2021nhg,Altmannshofer:2021qrr,Hurth:2021nsi,Carvunis:2021jga}. Most of these analyses suggested new physics solutions in the form of vector  and axial-vector operators. However there is no unique solution. In the simplest approach, where  only one new physics Wilson coefficient  or two related new physics Wilson coefficients are considered, the following scenarios provide a good fit to all $ b\to s\, \mu^+ \, \mu^- $ data:
\begin{itemize}
\item Scenario I:  In this scenario, the new physics is in the form of the operator $O_9=(\bar{s} \gamma^\mu P_L b)\, (\bar{\mu} \gamma^\mu \mu)$ alone. Its Wilson coefficient is $C_9 = C_9^{\rm SM} + C_9^{\rm NP}$ and the data require a large negative 
value of the NP  Wilson coefficient $C_9^{\rm NP}$.
\item Scenario II: The NP operators of this scenario are a linear combination of $O_9$ and $O_{10}=(\bar{s} \gamma^\mu P_L b)\, (\bar{\mu} \gamma^\mu \gamma^5 \mu)$. The Wilson coefficient of the latter operator is  $C_{10} =  C_{10}^{\rm SM} +
C_{10}^{\rm NP}$. The data imposes the condition $C_9^{\rm NP} = -C_{10}^{\rm NP}$ on the NP Wilson coefficients.
\item Scenario III: This scenario contains NP as a linear combination of $O_9$ and a non-SM operator $O'_9=(\bar{s} \gamma^\mu P_R b)\, (\bar{\mu} \gamma^\mu \mu)$ (the chirality flipped counterpart of $O_9$). A good fit to the data is achieved with  $C_9^{\rm NP} = -C_9^{\rm 'NP}$, where  $C_9^{\rm 'NP}$ is the Wilson coefficient of the operator $O'_{9}$.
\end{itemize}

Therefore one of the key open problems is to uniquely identify the Lorentz structure of new physics in $ b\to s\mu^+ \mu^- $  decay. It requires the development of techniques to discriminate between various possible solutions. These techniques may involve
\begin{itemize}
\item observing new decay modes driven by $b \to s \mu^+ \mu^-$~\cite{Grinstein:2015aua,Kumar:2017xgl,Kumbhakar:2018uty,Guadagnoli:2017quo,Abbas:2018xdu,Amhis:2020phx},
\item constructing new observables in the existing decay modes~\cite{Egede:2008uy,Egede:2010zc,Capdevila:2016ivx,Alguero:2019pjc} and
\item improving the precision in the present measurements.
\end{itemize}
In this work we show that a precision measurement of the branching ratio of the decay $B_s\to \mu^+\mu^-$ can lead to a clear distinction between scenario II and the other two scenarios. We also find  that the angular observables in the decay $B\rightarrow K^* \mu^+ \mu^-$, dependent on the azimuthal angle $\phi$, enable us to make a distinction between scenarios I and III, provided they
can be measured with small enough uncertainties. 
 
The paper is organized as follows. In sec. II, we discuss our strategies followed by three subsections. 
In subsection A, we show that scenario II predicts a much lower branching ratio for the decay $B_s\to \mu^+\mu^-$ 
compared to the other two scenarios. In subsection B, we obtain predictions for various azimuthal angular observables 
in $B\to K^*\mu^+\mu^-$  for the SM as well as for the allowed new physics scenarios. Further, we discuss the ability 
of these observables to discriminate between different NP solutions and show that the $S_9$ can
distinguish scenario III from scenario I, provided it can be measured with small enough uncertainty. 
In subsection C, we define azimuthal angle asymmetry $A_9$, proportional $S_9$, which can be measured with the smallest statistical uncertainty possible.
In sec. III, we present our conclusions.  

\section{Discrimination Variables}
In the SM, the effective Hamiltonian for $ b\to s \mu^+  \mu^- $ transition can be written as
\begin{eqnarray}
\mathcal{H}_{\rm SM} &=& - \frac{\alpha_{em} G_F}{\sqrt{2} \pi} V_{ts}^* V_{tb} \left[ 2 \frac{C_7^{\rm eff}}{q^2}
 [\overline{s} \sigma^{\mu \nu} q_\nu (m_s P_L  + m_b P_R)b] \bar{\mu} \gamma_\mu \mu \right. \nonumber \\ 
& & \left. + C_9^{\rm eff} (\overline{s} \gamma^{\mu} P_L b)(\overline{\mu} \gamma_{\mu} \mu) + C_{10} (\overline{s} \gamma^{\mu} P_L b)(\overline{\mu} \gamma_{\mu} \gamma_5 \mu) \right] + h.c. ,
\end{eqnarray} 
where $\alpha_{em}$ is the fine-structure constant, $G_F$ is the Fermi constant, $V_{ts}$ and $V_{tb}$ are the Cabibbo-Kobayashi-Maskawa (CKM) matrix elements and $P_{L,R} = (1 \mp \gamma^{5})/2$ are the chiral projection operators. The $q$ in the
$C_7$ term is the momentum of the off-shell photon in the effective $b \to s \gamma^*$ transition. 

The new physics solutions which can explain all the $ b\rightarrow s \mu^+ \mu^-$  data are only in the form of vector  and axial-vector  operators. Hence we consider the addition of only these  operators to the SM Hamiltonian for both left and right chiral quark currents. Therefore, the new physics effective Hamiltonian for $b\rightarrow s\mu^+\mu^-$ process takes the form
\begin{eqnarray}
\mathcal{H}_{\rm NP} &=& -\frac{\alpha_{\rm em} G_F}{\sqrt{2} \pi} V_{ts}^* V_{tb} \left[ C^{\rm NP}_9 (\overline{s} \gamma^{\mu} P_L b)(\overline{\mu} \gamma_{\mu} \mu) + C^{\rm NP}_{10} (\overline{s} \gamma^{\mu} P_L b)(\overline{\mu} \gamma_{\mu} \gamma_5 \mu) \right. \nonumber \\
& & \left. + C^{\prime \rm NP}_9 (\overline{s} \gamma^{\mu} P_R b)(\overline{\mu} \gamma_{\mu} \mu) + C^{\prime \rm NP}_{10} (\overline{s} \gamma^{\mu} P_R b)(\overline{\mu} \gamma_{\mu} \gamma_5 \mu)\right] + h.c.  ,
\label{HNP}
\end{eqnarray} 
where $C^{\rm NP}_{9,10}$ and $C^{\prime\rm NP}_{9,10}$ are the new physics Wilson coefficients. These  Wilson coefficients have been determined by a global fit to the all $ b\rightarrow s \mu^+ \mu^-$ data by different groups. A common conclusion of these global fits is that there are  three new physics solutions to  $ b\rightarrow s \mu^+ \mu^-$ 
anomalies\footnote{There can be other new physics scenarios, such as $C_{10}^{NP}$ and $C_{9}^{NP}=C'_{10}$, providing a good fit to the data \cite{Alguero:2019ptt}. However, $\Delta \chi^2=\chi^2_{\rm SM} - \chi^2_{\rm NP}$  for these solutions are smaller in comparison to scenarios I, II and III for which $\Delta \chi^2 \geq 44$. On the other hand, $\Delta \chi^2$ for  $C_{10}^{NP}$ and $C_{9}^{NP}=C'_{10}$ scenarios are $~34$ and $28$, respectively. Therefore we do not consider these moderate solutions in our analysis.}. 
These scenarios along with the fit values of Wilson coefficients are listed in Table~\ref{tab1}.
\begin{table}[htbp]
\centering
\tabcolsep 10pt
\begin{tabular}{|c|c|c|}
\hline\hline
NP scenarios & Best fit value &  pull\\
\hline
(I) $C^{\rm NP}_{9}$ & $-1.01\pm 0.15$ & 6.9\\
\hline
(II) $C^{\rm NP}_9 = -C^{\rm NP}_{10}$  & $-0.49\pm 0.07$ & 7.0 \\
\hline
 (III) $C^{\rm NP}_9 = -C^{\prime \rm NP}_{9}$ & $-1.03\pm 0.15$ & 6.7  \\
\hline\hline
\end{tabular}
\caption{The best fit values of the Wilson coefficients and the corresponding pull values are calculated using the methodology of Ref.~\cite{Alok:2019ufo} after Moriond 2021. Here pull value = $\sqrt{\chi^2_{\rm SM} - \chi^2_{\rm NP}}$.}
\label{tab1}
\end{table}

In the following subsections, we discuss methods to distinguish between these solutions by investigating $B_s\to \mu^+\mu^-$ and $B\to K^*\mu^+\mu^-$ decays. The angular observables in $B\to K^*\mu^+\mu^-$ decay could be standard tools to discriminate the NP solutions. In Refs.~\cite{Alok:2016qyh,Bhattacharya:2018kig,Alok:2018uft,Huang:2018nnq,Alok:2019uqc,Murgui:2019czp,Shi:2019gxi,Blanke:2019qrx,Asadi:2019xrc}, it is shown that the longitudinal polarization fraction of the vector meson and the forward-backward asymmetry can only discriminate the tensor and scalar NP solutions. Hence these two observables could not help us. Therefore, we look for those observables which depend on the azimuthal angle of the $B\to K^*\mu^+\mu^-$ decay .

\subsection{Distinguishing power of $B_s\to \mu^+\mu^-$}

The amplitude for the decay $B_s\to \mu^+\mu^-$ is non-zero only when both the quark and the lepton bi-linears are of axial vector form. All four NP operators contain quark axial vector current but only $O_{10}$ and $O_{10}^\prime$ contain the
lepton axial current. Hence only these two operators contribute to this decay.
In the presence of the NP Hamiltonian of Eq.~(\ref{HNP}), the matrix element can be written as
\begin{equation}
i\mathcal{M}_{B_s\to \mu\mu} = -\frac{i}{2}\frac{4G_F}{\sqrt{2}}\frac{\alpha_{\rm em}}{4\pi} V_{tb}V^*_{ts} (C_{10}+C^{\rm NP}_{10}-C^{\prime {\rm NP}}_{10}) \langle 0|\bar{s}\gamma_{\alpha}\gamma_5b| B_s(p)\rangle \left(\bar{\mu}\gamma^{\alpha}\gamma_5\mu\right).
\end{equation}
The corresponding hadronic matrix element is expressed as
\begin{equation}
\langle 0|\bar{s}\gamma_{\alpha}\gamma_5b| B_s(p)\rangle = i p_{\alpha} f_{B_s},
\end{equation} 
where $f_{B_s} = (230.3\pm 1.3)$ MeV~\cite{Aoki:2019cca} is the decay constant of $B_s$ meson. Therefore, the expression for the branching fraction is
\begin{equation}
\mathcal{B}(B_s\to \mu^+\mu^-) = \frac{G^2_F\alpha^2_{\rm em} m_{B_s} f^2_{B_s} m_\mu^2 \tau_{B_s}}{16\pi^3} |V_{tb}V^*_{ts}|^2\sqrt{1-\frac{4m^2_{\mu}}{m^2_{B_s}}} \left\vert (C_{10}+C^{\rm NP}_{10}-C^{\prime {\rm NP}}_{10}) \right\vert^2,
\label{br}
\end{equation} 
where $\tau_{B_s} = (1.527\pm 0.011)$ ps is the lifetime of the $B_s$ meson~\cite{Tanabashi:2018oca}. 

The SM prediction of this quantity is $\mathcal{B}(B_s\to \mu^+ \mu^-)|_{\rm SM} = (3.66\pm 0.16)\times 10^{-9}$~\cite{Straub:2018kue} which includes the QED corrections and agrees with the prediction of Ref.~\cite{Beneke:2019slt}. From the expression of Eq.~(\ref{br}), it is evident that $\mathcal{B}(B_s\to \mu^+\mu^-)$ is affected only by the NP Wilson coefficients 
$C_{10}^{\rm NP}$ and $C_{10}^{\prime {\rm NP}}$. Of the three NP scenarios allowed by the data, only the NP scenario II
contributes to this decay. For this scenario, the predicted value of the branching ratio is
\begin{equation}
\mathcal{B}(B_s\to \mu^+\mu^-)|_{\rm S\,II}= (2.77\pm 0.12)\times 10^{-9},
\end{equation} 
whereas the other two NP scenarios predict it to be the same as the SM value. 
The present experimental average of this branching fraction is~\cite{Altmannshofer:2021qrr}
\begin{equation}
\mathcal{B}(B_s\to \mu^+ \mu^-)|_{\rm exp}= (2.93\pm 0.35)\times 10^{-9}.
\end{equation}

The experimental central value is closer to the prediction of scenario II, compared to the other two scenarios. However, the present experimental uncertainty is reasonably large and we can not make a discrimination between scenario II and the other two scenarios. If a future measurement yields a value close to the prediction of scenario II, with an experimental uncertainty comparable to the theoretical uncertainty, then scenarios I and III are strongly disfavored. Such a reduction in experimental uncertainty is expected to be achieved at the end of Run-3 of LHC which will provide an integrated luminosity of $300$ fb$^{-1}$~\cite{Cerri:2018ypt,Bediaga:2018lhg}.

\subsection{Distinguishing through azimuthal angular asymmetries of $B\to K^*\mu^+\mu^-$}

To make a distinction between scenario I and scenario III, we turn to angular variables other than 
longitudinal polarization fraction of $K^*$ or the forward-backward asymmetry. 
The differential distribution of four-body decay $B\to K^*(\to K\pi)\mu^+\mu^-$ can be parametrized 
as the function of one kinematic and three angular variables. The kinematic variable is $q^2 = (p_B-p_{K^*})^2$, where $p_B$ and $p_{K^*}$ are respective four-momenta of $B$ and $K^*$ mesons. The angular variables are defined in the $K^*$ rest frame. They are (a) $\theta_{K}$ the angle between 
$B$ and $K$ mesons where $K$ meson comes from $K^*$ decay, (b) $\theta_{\mu}$ the angle between momenta of $\mu^-$ and $B$ meson and (c) $\phi$ the angle between $K^*$ decay plane and the plane defined by the $\mu^+-\mu^-$ momenta. The full decay distribution can be expressed as~\cite{Bobeth:2008ij, Altmannshofer:2008dz}
\begin{equation}
\frac{d^4\Gamma}{dq^2d\cos\theta_{\mu}d\cos\theta_{K}d\phi} = \frac{9}{32\pi}I(q^2,\theta_{\mu},\theta_{K},\phi),
\end{equation}
where
\begin{eqnarray}
I(q^2,\theta_{\mu},\theta_{K},\phi) &=& I^s_1\sin^2\theta_{K} + I^c_1\cos^2\theta_{K}+(I^s_2\sin^2\theta_{K}+I^c_2\cos^2\theta_{K})\cos 2\theta_{\mu} \nonumber\\
& & +I_3\sin^2\theta_{K}\sin^2\theta_{\mu}\cos 2\phi +I_4\sin 2\theta_{K}\sin 2\theta_{\mu}\cos\phi \nonumber\\
& & + I_5 \sin 2\theta_{K}\sin\theta_{\mu}\cos\phi \nonumber \\
& & + (I^s_6\sin^2\theta_{K}+I^c_6\cos^2\theta_{K})\cos\theta_{\mu} + I_7\sin 2\theta_{K}\sin\theta_{\mu}\sin\phi \nonumber\\
& & + I_8\sin 2\theta_{K}\sin 2 \theta_{\mu} \sin\phi +I_9\sin^2\theta_{K}\sin^2\theta_{\mu}\sin 2\phi .
\label{Ifunc}
\end{eqnarray}
The twelve angular coefficients $I^{(a)}_i$ depend on $q^2$ and on various hadron form factors.
The detailed expressions of these coefficients are given in Appendix~(\ref{appen}). The corresponding expression for the CP conjugate of the decay can be obtained by replacing $\theta_{\mu}$ by $(\pi-\theta_{\mu})$ and $\phi$ by $-\phi$. This leads to the following transformations of angular coefficients
\begin{equation}
I^{(a)}_{1,2,3,4,7} \Longrightarrow \bar{I}^{(a)}_{1,2,3,4,7}, \quad I^{(a)}_{5,6,8,9} \Longrightarrow -\bar{I}^{(a)}_{5,6,8,9},
\end{equation}
where $\bar{I}^{(a)}_i$ are the complex conjugate of $I^{(a)}_i$. Therefore, there could be twelve CP averaged angular observables which can be defined as~\cite{Bobeth:2008ij,Altmannshofer:2008dz}
\begin{equation}
S^{(a)}_i(q^2) = \frac{I^{(a)}_i(q^2)+ \bar{I}^{(a)}_i(q^2)}{d(\Gamma +\bar{\Gamma})/dq^2}.
\end{equation}

\begin{figure}[htbp]
\begin{tabular}{cc}
\includegraphics[width=88mm]{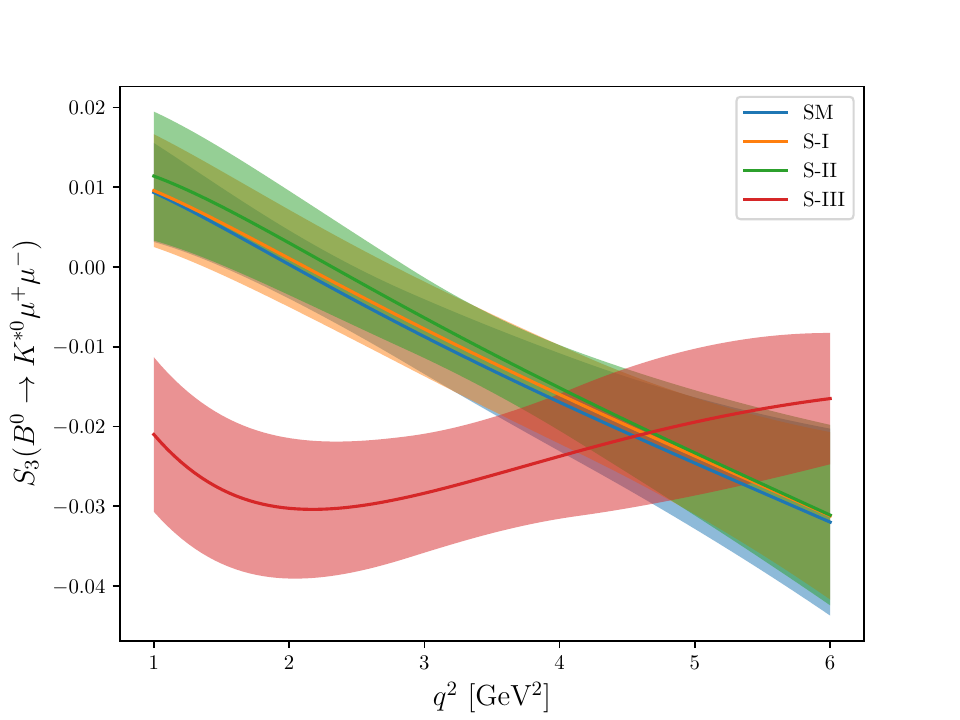} & \includegraphics[width=88mm]{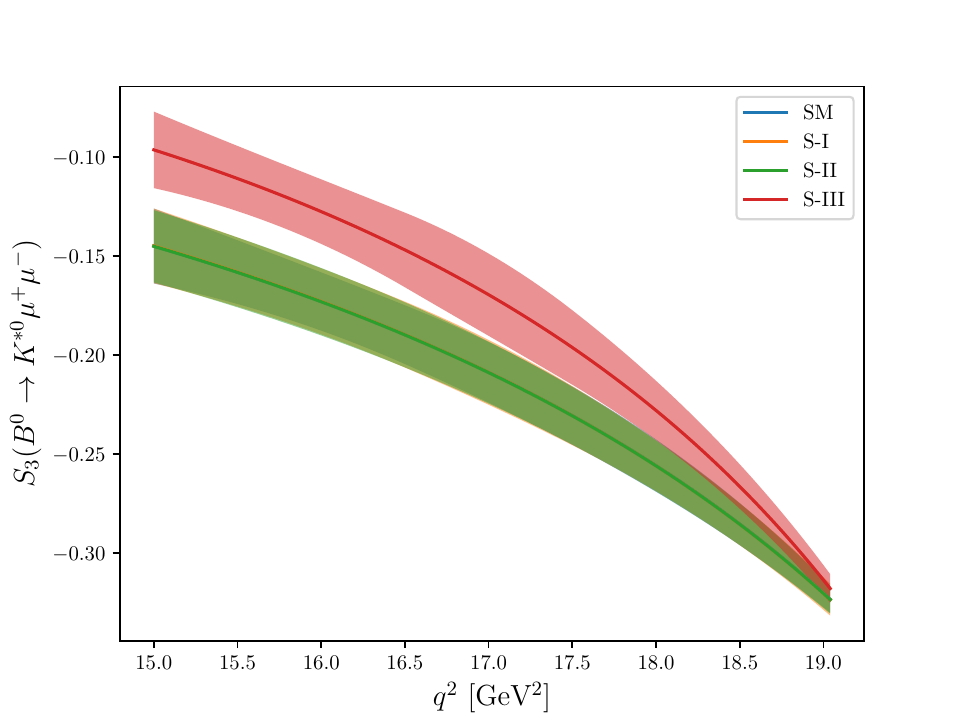}\\
\includegraphics[width=88mm]{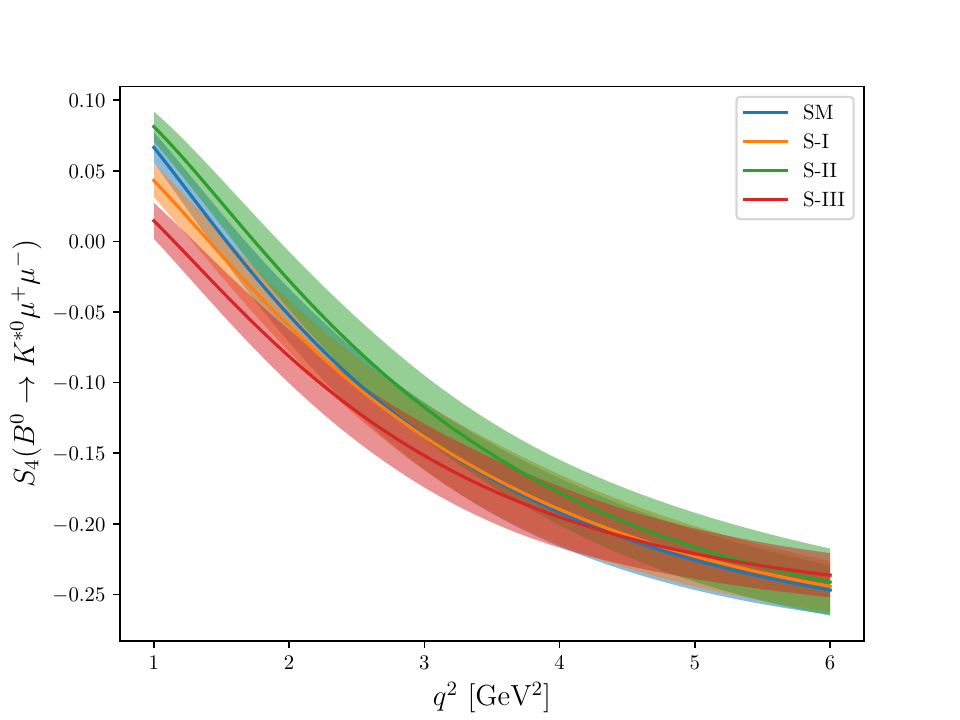} & \includegraphics[width=88mm]{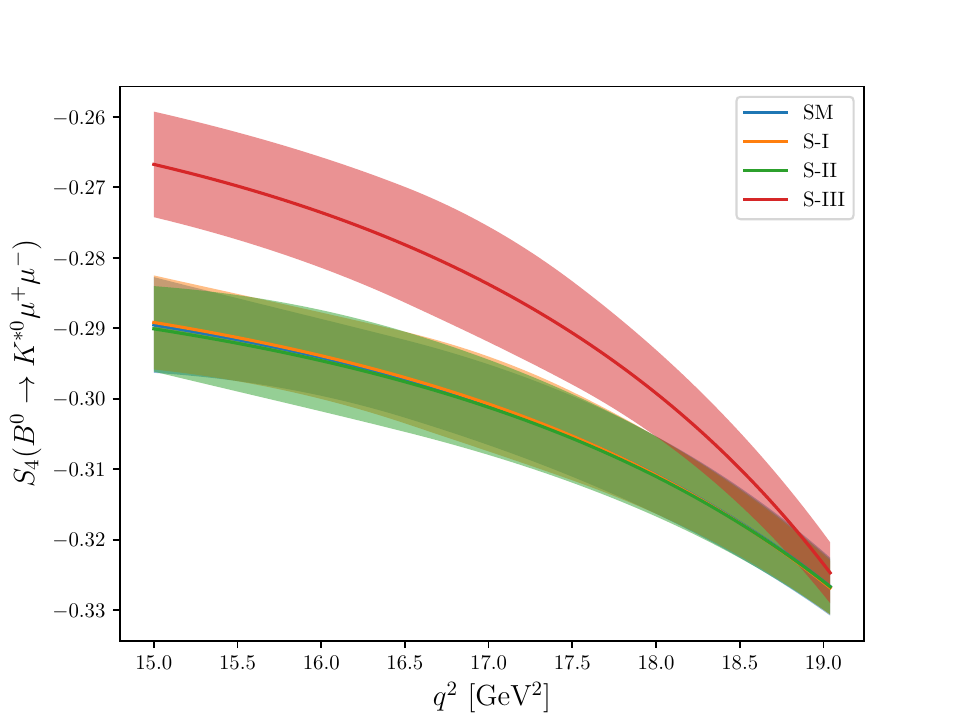}\\
\includegraphics[width=88mm]{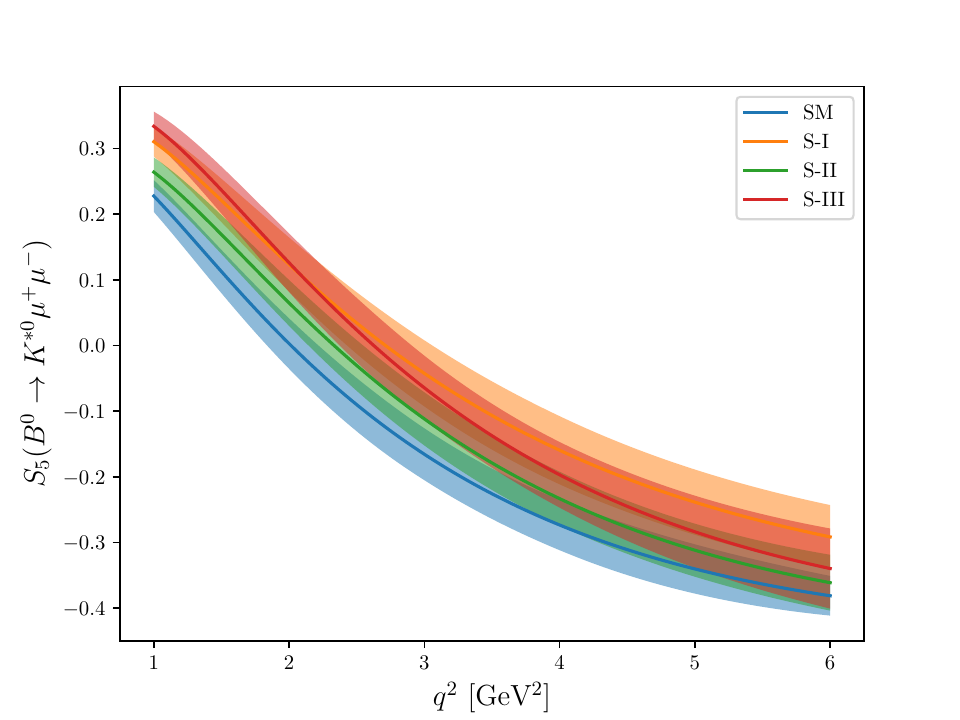} & \includegraphics[width=88mm]{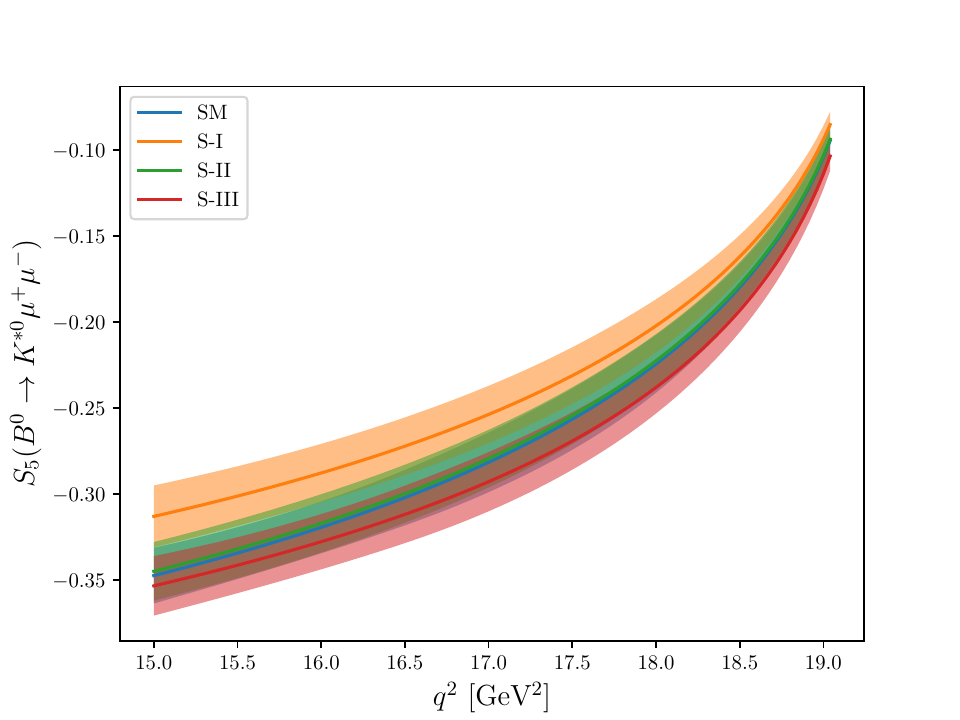}\\
\end{tabular}
\caption{Plots of $S_{3,4,5}(q^2)$ as a function of $q^2$ for SM and three NP scenarios. The left and right panels correspond to the low ($[1.1,6.0]$ GeV$^2$) and high ($[15,19]$ GeV$^2$) $q^2$ bins respectively. In each plot, the band represents the theoretical uncertainty mainly due to the form factors. Note that the scale on the y axis on each plot is different.}
\label{fig1}
\end{figure}

\begin{figure}[htbp]
\begin{tabular}{cc}
\includegraphics[width=88mm]{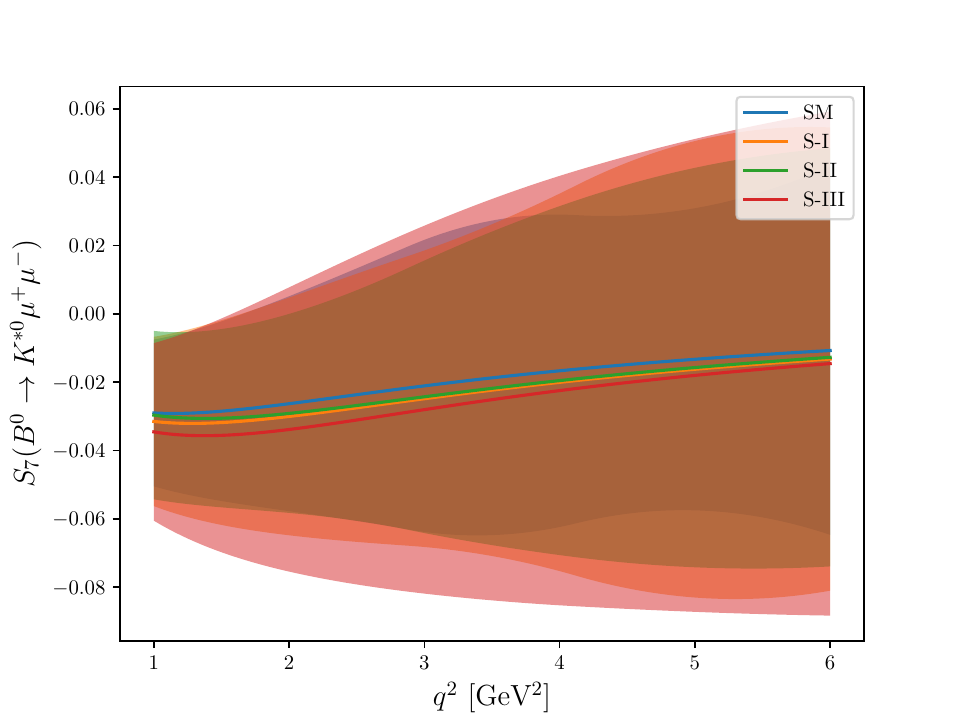} & \includegraphics[width=88mm]{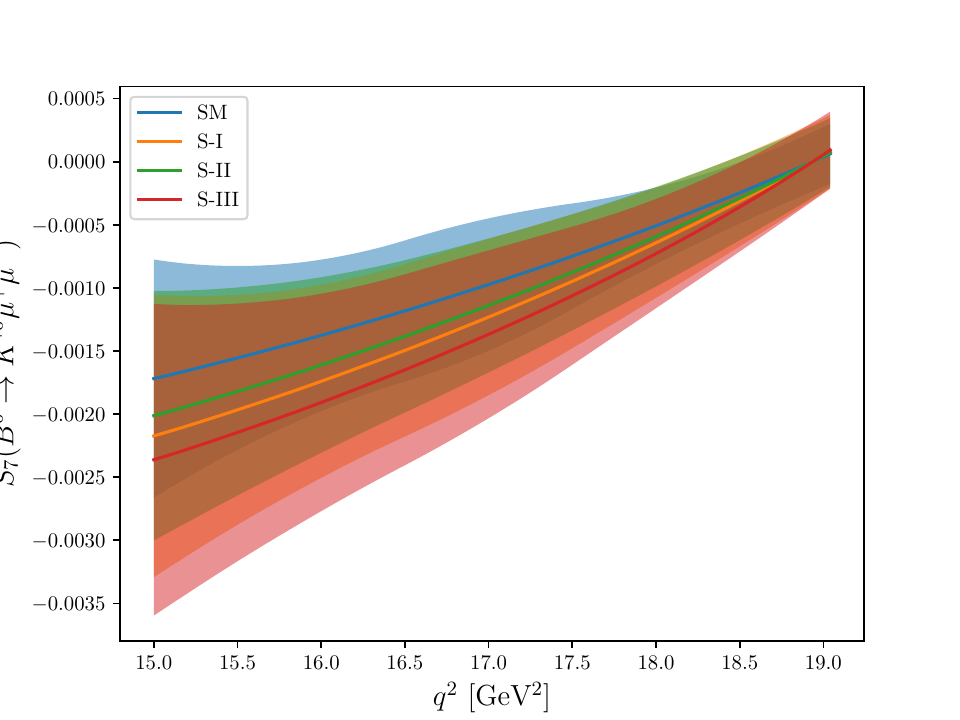}\\
\includegraphics[width=88mm]{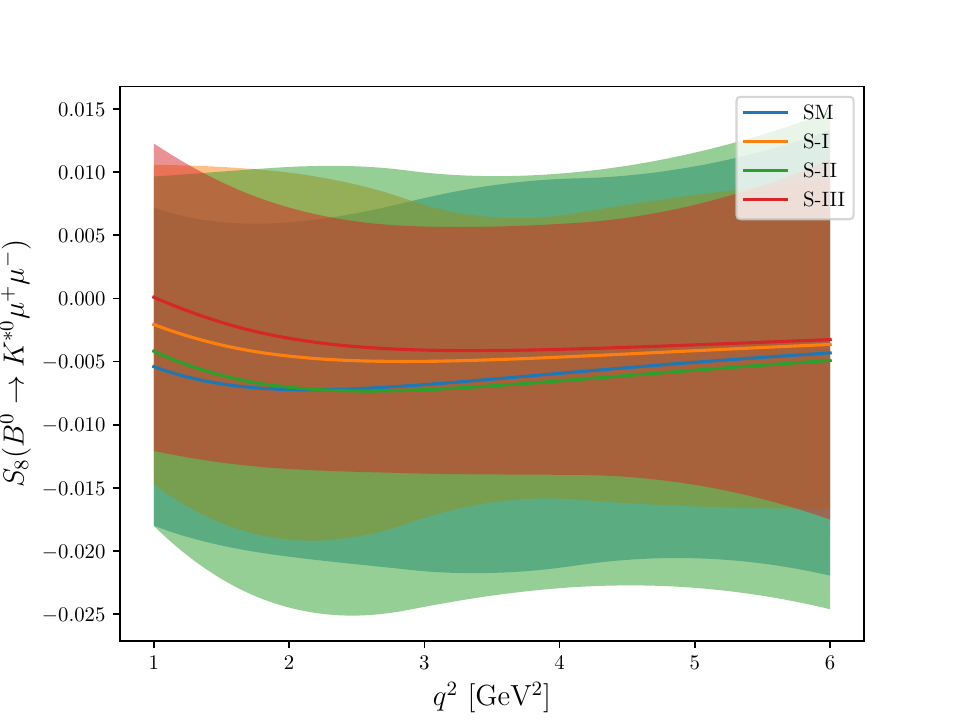} & \includegraphics[width=88mm]{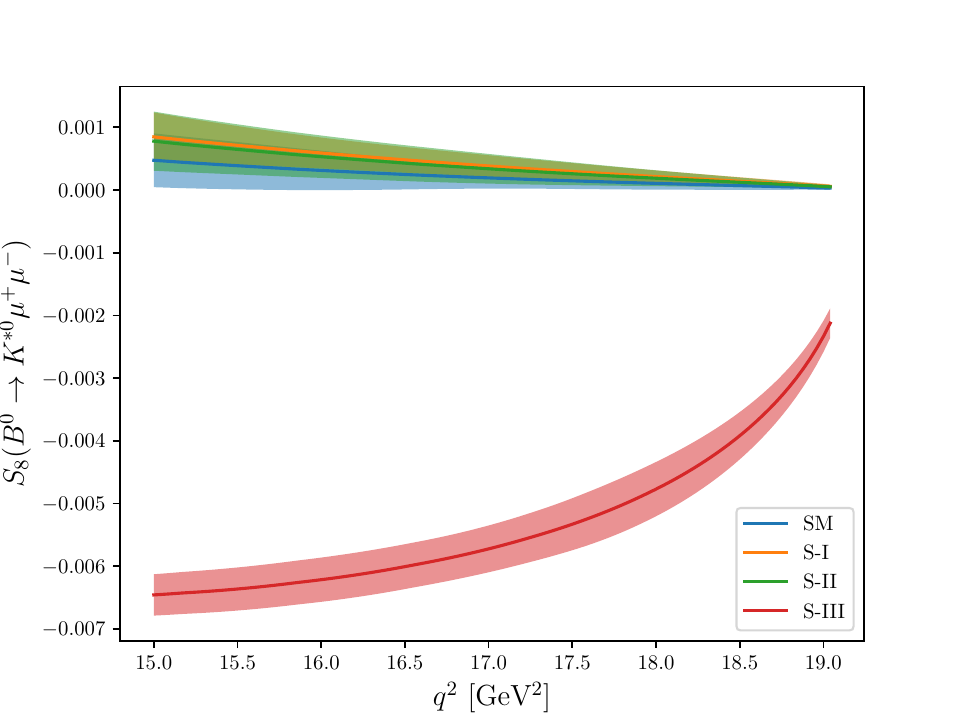}\\
\includegraphics[width=88mm]{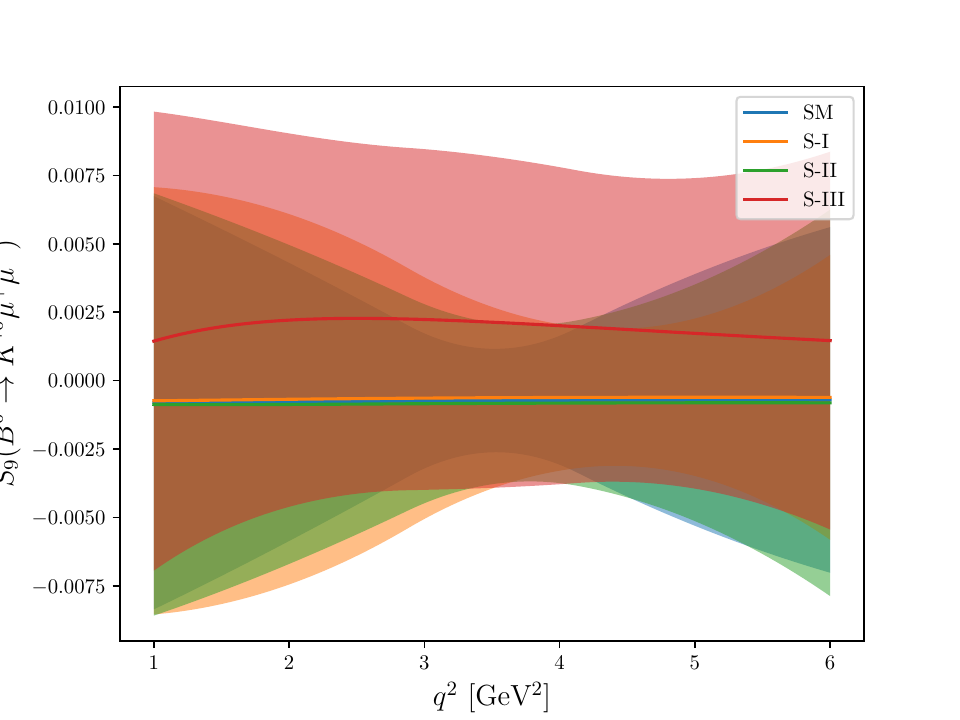} & \includegraphics[width=88mm]{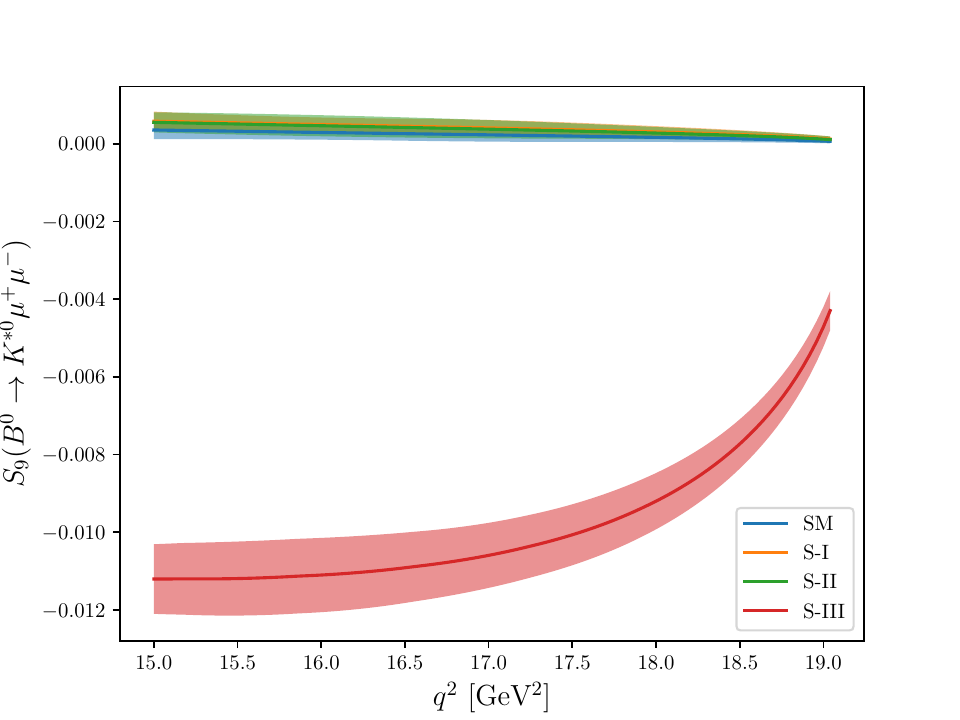}\\
\end{tabular}
\caption{Plots of $S_{7,8,9}(q^2)$ as a function of $q^2$ for SM and three NP scenarios. The left and right panels correspond to the low ($[1.1,6.0]$ GeV$^2$) and high ($[15,19]$ GeV$^2$) $q^2$ bins respectively. In each plot, the band represents the theoretical uncertainty mainly due to the form factors. Note that the scale on the y axis on each plot is different.}
\label{fig2}
\end{figure}

The longitudinal polarization fraction of $K^*$ depends on the distribution of the events in the angle $\theta_K$ (after
integrating over $\theta_\mu$ and $\phi$) and the forward-backward asymmetry is defined in terms of $\theta_\mu$
(after integrating over $\theta_K$ and $\phi$). Both these quantities have very poor discrimination for NP other than
scalar or tensor operators. Therefore, we study the observables that are based on the distribution in the azimuthal angle $\phi$. 
In particular, we investigate the distinguishing ability of $S_{3,4,5}$ and $S_{7,8,9}$. We compute the average values of these six observables for the SM and the three NP scenarios in four different $q^2$ bins, $q^2\subset [1.1,6.0],\, [15,17],\, [17,19]$ and $[15,19]$ GeV$^2$. These are listed in Tab~\ref{tab2}. In this table, we also mention current measured values of these six quantities. We plot the six observables as a function of $q^2$ for the SM and the three NP scenarios. The $q^2$ plots for $S_{3,4,5}(q^2)$ are shown in Fig.~\ref{fig1} whereas those for $S_{7,8,9}(q^2)$ are given in Fig.~\ref{fig2}. The average values and the plots are obtained by using {\tt Flavio} package~\cite{Straub:2018kue}. This package uses the most precise form factor predictions obtained in light cone sum rule 
(LCSR)~\cite{Straub:2015ica, Gubernari:2018wyi} approach, taking into account the correlations between the uncertainties of different form factors and at different values of $q^2$. The non-factorisable corrections are incorporated following the parameterization used in Ref.~\cite{Straub:2015ica,Straub:2018kue}. These are also compatible with the calculations in Ref.~\cite{ Khodjamirian:2010vf}.

\begin{table}[ht]
\begin{tabular}{c|c|c|c|c|c|c}
\hline
Observable & $q^2$ bin & SM & S-I & S-II & S-III & Expt. value ({\tt LHCb}) \\
\hline
 & $[1.1, 6]$ & $-0.013\pm 0.005$ & $-0.012\pm 0.005$ & $-0.011\pm 0.005$ & $-0.027\pm 0.007$ & $-0.012 \pm 0.025\pm 0.003$\\
$S_3$ & $[15,17]$ & $-0.173\pm 0.019$ & $-0.173\pm 0.017$ & $-0.173\pm 0.016$ & $-0.124\pm 0.019$ & $-0.166\pm 0.034\pm  0.007$\\
 & $[17,19]$ & $-0.251\pm 0.013$ & $-0.252\pm 0.012$ & $-0.252\pm 0.013$& $-0.220\pm 0.016$  & $-0.250\pm 0.050\pm 0.025$ \\
 & $[15,19]$ & $-0.205\pm 0.015$ & $-0.205\pm 0.016$ & $-0.205\pm 0.014$ & $-0.162\pm 0.018$ & $-0.189 \pm 0.030\pm 0.009$\\
 \hline
 & $[1.1, 6]$ & $-0.147\pm 0.019$ & $-0.147\pm 0.020$ & $-0.130\pm 0.020$ & $-0.159\pm 0.017$ & $-0.136\pm 0.039\pm 0.003$\\
$S_4$ & $[15,17]$ & $-0.294\pm 0.006$ & $-0.294\pm 0.006$ & $-0.294\pm 0.007$ & $-0.272\pm 0.007$ & $-0.299\pm 0.033\pm 0.008$\\
 & $[17,19]$ & $-0.310\pm 0.006$ & $-0.310\pm 0.006$ & $-0.310\pm 0.006$ & $-0.297\pm 0.006$ & $-0.307\pm 0.041\pm 0.008$\\
 & $[15,19]$ & $-0.300\pm 0.006$ & $-0.300\pm 0.007$ & $-0.300\pm 0.006$ & $-0.282\pm 0.007$ & $-0.303\pm 0.024\pm 0.008$\\
 
 \hline
  & $[1.1, 6]$ & $-0.186\pm 0.037$ &  $-0.074\pm 0.046$ & $-0.142\pm 0.042$ & $-0.081\pm 0.055$ & $-0.052\pm 0.034\pm 0.007$\\
$S_5$ & $[15,17]$ & $-0.318\pm 0.015$ & $-0.288\pm 0.015$ & $-0.316\pm 0.016$ & $-0.324\pm 0.016$ & $-0.341\pm 0.034\pm 0.009$\\
 & $[17,19]$ & $-0.226\pm 0.017$ & $-0.205\pm 0.016$ & $-0.224\pm 0.015$ & $-0.237\pm 0.016$ & $-0.280\pm 0.040\pm 0.014$\\
 & $[15,19]$ & $-0.280\pm 0.017$ & $-0.254\pm 0.017$ & $-0.278\pm 0.016$ & $-0.289\pm 0.016$ & $-0.317\pm 0.024\pm 0.011$ \\
\hline
 & $[1.1, 6]$ & $-0.019\pm 0.041$ & $-0.023\pm 0.042$ & $-0.022\pm 0.041$ & $-0.025\pm 0.046$ & $-0.090\pm 0.034\pm 0.002$\\
$S_7$ & $[15,17]$ & $-0.001\pm 0.001$ & $-0.002\pm 0.001$ & $-0.002\pm 0.001$ & $-0.002\pm 0.001$ & $0.029\pm 0.039\pm 0.001$\\
 & $[17,19]$ & $-0.001\pm 0.001$ & $-0.001\pm 0.000$ & $-0.001\pm 0.000$ & $-0.001\pm 0.001$ & $0.049\pm 0.049\pm 0.007$\\
 & $[15,19]$ & $-0.001\pm 0.001$ & $-0.001\pm 0.001$ & $-0.001\pm 0.001$ & $-0.001\pm 0.001$ & $0.035\pm 0.030\pm 0.003$\\
 \hline
  & $[1.1, 6]$ & $-0.006\pm 0.014$ & $-0.004\pm 0.013$ & $-0.006\pm 0.015$ & $-0.003\pm 0.007$ & $-0.009\pm 0.037\pm 0.002$\\
$S_8$ & $[15,17]$ & $0.000\pm 0.000$ & $0.000\pm 0.000$ & $0.000\pm 0.000$ & $-0.006\pm 0.001$ & $0.003\pm 0.042\pm 0.002$\\
 & $[17,19]$ & $0.000\pm 0.000$ & $0.000\pm 0.000$ & $0.000\pm 0.000$ & $-0.005\pm 0.001$ & $-0.026\pm 0.046\pm 0.002$ \\
 & $[15,19]$ & $0.000\pm 0.000$ & $0.000\pm 0.000$ & $0.000\pm 0.000$ & $-0.006\pm 0.001$ & $0.005\pm 0.031\pm 0.001$\\
 \hline
  & $[1.1, 6]$ & $-0.001\pm 0.002$ & $-0.001\pm 0.003$ & $-0.001\pm 0.003$ & $0.002\pm 0.005$ & $-0.025\pm 0.026\pm 0.002$\\
$S_9$ & $[15,17]$ & $0.000\pm 0.000$ & $0.001\pm 0.000$ & $0.000\pm 0.000$ & $-0.012\pm 0.001$ & $0.000\pm 0.037\pm 0.002$\\
 & $[17,19]$ & $0.000\pm 0.000$ & $0.000\pm 0.000$ & $0.000\pm 0.000$ & $-0.010\pm 0.001$ & $-0.056\pm 0.045\pm 0.002$\\
 & $[15,19]$ & $0.000\pm 0.000$ & $0.000\pm 0.000$ & $0.000\pm 0.000$ & $-0.012\pm 0.001$ & $-0.031\pm 0.029\pm 0.001$\\
 \hline
\end{tabular}
\caption{Average values of $S_{3,4,5}$ and $S_{7,8,9}$ in four $q^2$ bins for the SM and three NP scenarios listed in Tab~\ref{tab1}. Present experimental measurements of these quantities are also listed for comparison~\cite{Aaij:2020nrf}.}
\label{tab2}
\end{table}

From the Figs.~\ref{fig1}, \ref{fig2} and Tab~\ref{tab2}, we make following observations:
\begin{itemize}
\item
The values of $S_i^{(a)}$ in the low-$q^2$ bin are lower compared to the values in the high-$q^2$ bin. The values of the 
observables $S_{3,4,5,7,8,9}$ in the low-$q^2$ bin do not have any ability to discriminate between three NP scenarios. 

\item 
In high $q^2$ bin, the $S_{5}$ and $S_7$ do not have any kind of discrimination power, whereas $S_4$ has a poor distinguishing capability for the NP scenario III. In addition, $S_8$ can also discriminate the third scenario, but the average values are less than 
$1\%$. Therefore, $S_4$ and $S_8$ are poor distinguishing tools.
\item 
The prediction of NP scenario III and that of NP scenario I, for $S_3$ in high-$q^2$ bin, differ from each other by 
about $20\%$. But these predictions have a theoretical uncertainty of $10\%$. This observable becomes an effective 
distinguishing tool if the theoretical uncertainty can be reduced to $5\%$ and if the experimental uncertainty can also be reduced to a similar level.

\item It is advantageous to use $S_9$ as a discriminator for NP scenario III because its theoretical uncertainty is negligibly small. NP scenario III predicts the value of $S_9$ in the high $q^2$ bin to be about a percent, whereas the predictions of the other two NP scenarios are zero. Measuring $S_9$ to a precision of $0.5\%$ leads to a $1\sigma$ distinction between NP scenario III from the other two. For $2\sigma$ distinction, the experimental uncertainty should be reduced by an additional factor of 2.
\end{itemize}

\subsection{Measurement of $S_3$ and $S_9$ with the smallest possible uncertainty}

The number of $B \to K^* \mu^+ \mu^-$ events in an experiment are likely to be limited
because of the very small branching ratio. If this small set of events is fitted to the full differential distribution
in $q^2$ as well as in all the three angles $\theta_K$, $\theta_\mu$ and $\phi$ to determine $S_i^{(a)}(q^2)$, 
the number of events in each bin will be rather small and the statistical uncertainties in such a determination 
will be quite large. It is possible to improve the statistics, by integrating over the polar angles $\theta_K$ and 
$\theta_\mu$ \cite{Bobeth:2008ij} and define the two distributions
\begin{eqnarray}
I_{\rm sum} (q^2, \phi) & = & \int_{-1}^1 d \cos \theta_K \int_{-1}^1 d \cos \theta_\mu \left[ I (q^2, \theta_K, \theta_\mu, \phi) 
+ \bar{I} (q^2, \theta_K, \theta_\mu, \phi) \right]\,  \nonumber \\
I_{\rm diff} (q^2, \phi) & = & \int_{-1}^1 d \cos \theta_K \int_{-1}^1 d \cos \theta_\mu \left[ I (q^2, \theta_K, \theta_\mu, \phi) 
- \bar{I} (q^2, \theta_K, \theta_\mu, \phi) \right].
\label{IsumIdiff}
\end{eqnarray}
By doing a fit of $I_{\rm sum}(q^2,\phi)$ and $I_{\rm diff}(q^2,\phi)$ data binned in angle $\phi$,
it is possible to determine the coefficient of $\cos 2 \phi$ $(S_3)$ and of $\sin 2 \phi$ $(S_9)$. However, it also is possible to 
measure $S_3$ and $S_9$ by considering $I_{\rm sum}(q^2,\phi)$ and $I_{\rm diff}(q^2,\phi)$ in $90^{\circ}$ wide bins of $\phi$ and define the two asymmetries
\begin{equation}
A_3 (q^2)= \frac{\left(\int^{\pi/4}_{-\pi/4}-\int^{3\pi/4}_{\pi/4}+ \int^{5\pi/4}_{3\pi/4}-\int^{7\pi/4}_{5\pi/4}\right) 
d \phi I_{\rm sum}(q^2, \phi)}{\left(\int^{\pi/4}_{-\pi/4}+\int^{3\pi/4}_{\pi/4}+ \int^{5\pi/4}_{3\pi/4}+
\int^{7\pi/4}_{5\pi/4}\right)d\phi I_{\rm sum} (q^2,\phi)},
\label{defA3}
\end{equation}
and \cite{Mandal:2014kma}
\begin{equation}
A_9 (q^2) = \frac{\left(\int^{\pi/2}_{0}-\int^{\pi}_{\pi/2}+ \int^{3\pi/2}_{\pi}-\int^{2\pi}_{3\pi/2}\right) 
d \phi I_{\rm diff} (q^2, \phi)}{\left(\int^{\pi/2}_{0}+\int^{\pi}_{\pi/2}+ \int^{3\pi/2}_{\pi}+\int^{2\pi}_{3\pi/2}\right)
d\phi I_{\rm sum} (q^2,\phi)}.
\label{defA9}
\end{equation}
It is straight forward to show that $A_3 = (2/\pi) S_3$ and $A_9 = (2/\pi) S_9$. 
Since $A_3$ and $A_9$ are defined using the largest possible bins in $\phi$, they can be measured with 
the least possible statistical uncertainty.
As discussed above, a determination of $A_9$, with low statistical error in the high $q^2$ bins, 
will lead to a clear distinction between the NP scenarios I and III.

\section{Conclusions}
The global fits of the current data on the semi-leptonic $b\to s$ transitions lead to three different NP solutions 
(I) $C_9^{\rm NP}<0$, (II) $C_{9}^{\rm NP} = -C_{10}^{\rm NP}$, (III) $C_9^{\rm NP} = -C_9^{\prime \rm NP}$. 
In this work, we suggest a method to uniquely determine which of these three solutions is the correct one by 
investigating $B_s\to \mu^+\mu^-$ and $B\to K^*\mu^+\mu^-$ decays. The $B_s\to \mu^+\mu^-$ amplitude 
is non-zero only if the leptonic current has an axial-vector component. Among the three solutions, only scenario II 
satisfies this constraint. Therefore, the branching ratio of this decay can distinguish scenario II from the other two, 
provided the present experimental uncertainty in its measurement is reduced by a factor of three. It is expected 
that the Run-3 of LHC will lead to such a precise measurement~\cite{Cerri:2018ypt}. To make a distinction between the other two scenarios, 
we study the azimuthal angular observables in the decay $B\to K^*\mu^+\mu^-$ and show that the observables $S_9$ in high $q^2$ bin is an effective tool to distinguish between the NP scenarios I and III, 
provided its uncertainty is small enough. We also define an asymmetry in the azimuthal angle $\phi$,  $A_9 = (2/\pi) S_9$. 
This is directly measurable and utilizes the largest possible bin sizes in $\phi$. So, for any given
data set, determination of $S_9$ through a measurement of $A_9$ leads to the smallest 
statistical uncertainty. Thus $A_9$ is a good tool to make a discrimination between scenarios I and III.

\section*{Acknowledgement}
We would like to thank Ulrik Egede for his useful comments on the first version of this work.

\section*{Data Availability Statement} This manuscript has no associated data. We have not used
any data file in this work which has to be deposited.

\appendix
\section{Angular coefficients}
\label{appen}
The angular coefficients in eq~(\ref{Ifunc}) can be expressed in terms of transversity amplitudes which are given by~\cite{Altmannshofer:2008dz}
\begin{eqnarray}
I_1^s &=& \frac{(2+\beta^2_{\mu})}{4}\left[|A^L_{\perp}|^2+|A^L_{\parallel}|^2 +(L\to R)\right] + \frac{4m^2_{\mu}}{q^2} {\rm Re}\left(A^L_{\perp}A^{R*}_{\perp}+A^L_{\parallel}A^{R*}_{\parallel}\right), \nonumber \\
I^c_1 & = & |A^L_{0}|^2+|A^R_{0}|^2 +\frac{4m^2_{\mu}}{q^2}\left[|A_t|^2 + 2 {\rm Re}\left(A^L_0 A^{R*}_0\right)\right] +\beta^2_{\mu} |A_S|^2,\nonumber \\
I_2^s &=& \frac{\beta^2_{\mu}}{4}\left[|A^L_{\perp}|^2+|A^L_{\parallel}|^2 + (L\to R)\right],\nonumber\\
I^c_2 &= & -\beta^2_{\mu} \left[|A^L_0|^2 + |A^R_0|^2\right], \nonumber \\
I_3 &= & \frac{\beta^2_{\mu}}{2} \left[|A^L_{\perp}|^2 - |A^L_{\parallel}|^2 + (L\to R)\right], \nonumber \\
I_4 &  = & \frac{\beta^2_{\mu}}{\sqrt{2}} \left[ {\rm Re}(A^L_0 A^{L*}_{\parallel}) + (L\to R)\right], \nonumber \\
I_5 &= &\sqrt{2} \beta_{\mu}\left[{\rm Re}(A^L_0 A^{L*}_{\perp}) -(L\to R)- \frac{m_{\mu}}{\sqrt{q^2}} {\rm Re}(A^L_{\parallel}A^*_S+A^R_{\parallel}A^*_S)\right],\nonumber \\
I^s_6 &= & 2\beta_{\mu}\left[{\rm Re}(A^L_{\parallel}A^{L*}_{\perp})- (L\to R)\right],\nonumber\\
I^c_6 &=& 4\beta_{\mu}\frac{m_{\mu}}{\sqrt{q^2}}{\rm Re}\left[A^L_0A^*_S + (L\to R)\right],\nonumber\\
I_7 & =& \sqrt{2}\beta_{\mu}\left[{\rm Im}(A^L_0 A^{L*}_{\parallel})- (L\to R) +\frac{m_{\mu}}{\sqrt{q^2}}{\rm Im}(A^L_{\perp}A^*_S + A^R_{\perp}A^*_S)\right],\nonumber\\
I_8 &= & \frac{\beta^2_{\mu}}{\sqrt{2}}\left[{\rm Im}(A^L_0A^{L*}_{\perp}) + (L\to R)\right],\nonumber \\
I_9 &= & \beta^2_{\mu}\left[{\rm Im}(A^{L*}_{\parallel}A^L_{\perp})+ (L\to R)\right].
\end{eqnarray}
The transversity amplitudes are written as
\begin{eqnarray}
A_{\perp L,R} &=& N\sqrt{2\lambda}\left[\left[(C^{\rm eff}_9 +C^{\rm eff\prime}_9)\mp (C_{10}+C^{\prime}_{10})\right]\frac{V(q^2)}{m_B+m_{K^*}}+\frac{2m_b}{q^2}C^{\rm eff}_7T_1(q^2)\right],\nonumber \\
A_{\parallel L,R} & =& -N\sqrt{2} (m^2_{B}-m^2_{K^*})\left[\left[(C^{ \rm eff}_9-C^{\rm eff \prime}_9)\mp (C_{10}-C^{\prime}_{10})\right]\frac{A_1(q^2)}{m_B-m_{K^*}}+\frac{2m_b}{q^2}C^{\rm eff}_7 T_2(q^2)\right], \nonumber \\
A_{0L,R} & =& -\frac{N}{2m_{K^*}\sqrt{q^2}}\left[ [(C^{\rm eff}_9 -C^{\rm eff \prime}_9)\mp (C_{10}-C^{\prime}_{10})] \right. \nonumber \\
& & \left\lbrace (m^2_B-m^2_{K^*}-q^2)(m_B+m_{K^*})A_1(q^2)-\lambda\frac{A_2(q^2)}{m_B +m_{K^*}}\right\rbrace \nonumber \\
&& \left. +2 m_b C^{\rm eff}_7 \left\lbrace(m^2_B+3m^2_{K^*}-q^2)T_2(q^2)-\frac{\lambda}{m^2_B-m^2_{K^*}}T_3(q^2)\right\rbrace\right], \nonumber \\
A_t & =& \frac{N}{\sqrt{q^2}}\sqrt{\lambda}\left[2(C_{10}-C^{\prime}_{10})+\frac{q^2}{m_{\mu}}(C_P-C^{\prime}_P)\right] A_0(q^2),\nonumber\\
A_S & = & -2N\sqrt{\lambda}(C_S-C^{\prime}_S)A_0(q^2),
\end{eqnarray}
where 
\begin{equation}
N= V_{tb}V^*_{ts}\left[\frac{G^2_F\alpha^2}{3.2^{10}\pi^5m^3_B}q^2\sqrt{\lambda}\beta_{\mu}\right]^{1/2},
\end{equation}
with $\lambda = m^4_B+m^4_{K^*}+q^4-2(m^2_B m^2_{K^*} +m^2_{B}q^2+m^{2}_{K^*}q^2)$ and $\beta_{\mu}= \sqrt{1-4m^2_{\mu}/q^2}$. The expressions of form-factors $V(q^2)$, $A_{0,1,2}(q^2)$ and $T_{1,2,3}(q^2)$ can be found in ref.~\cite{Straub:2015ica} which are calculated by a combined fit of Light Cone Sum Rule and lattice QCD approaches.

\end{document}